\documentclass[showpacs,twocolumn,preprintnumbers,prl,amssymb,floatfix]{revtex4-1}
\usepackage{graphicx}
\usepackage{dcolumn}
\usepackage{bm}
\usepackage{amsmath}
\usepackage{amsfonts}
\usepackage{amssymb}
\usepackage{ulem}
\usepackage{color}

\begin{document}
\title{Magnetoresistance and Magnetic Ordering Fingerprints in Hydrogenated Graphene}
\author{David Soriano$^{1,2}$}
\author{Nicolas Leconte$^{3}$}
\author{Pablo Ordej\'on$^{4}$}
\author{Jean-Christophe Charlier$^{3}$}
\author{Juan-Jose Palacios$^{5}$}
\author{Stephan Roche$^{6,7}$}
\affiliation{$^{1}$Departamento de F\'isica Aplicada, Universidad de Alicante, San Vicente del Raspeig, Alicante 03690, Spain}
\affiliation{$^{2}$ Instituto de Ciencia de Materiales de Madrid (CSIC), Cantoblanco, Madrid, 28049, Spain}
\affiliation{$^{3}$Universit\'e Catholique de Louvain, Institute of Condensed Matter and Nanoscience (IMCN), Place Croix du Sud 1 (NAPS-Boltzmann), 1348 Louvain-la-Neuve, Belgium}
\affiliation{$^{4}$CIN2 (CSIC-ICN) Barcelona, Campus UAB, E-08193 Bellaterra, Spain}
\affiliation{$^{5}$Depto. de F\'isica de la Materia Condensada, Universidad Aut\'onoma de Madrid, 
Cantoblanco, Madrid, 28049, Spain}
\affiliation{$^{6}$CIN2 (ICN-CSIC) and Universitat Autonoma de Barcelona, Catalan Institute of Nanotechnology, Campus de la UAB, 08193 Bellaterra (Barcelona), Spain}
\affiliation{$^{7}$ ICREA, Institucio Catalana de Recerca i Estudis Avan\c cats, 08010 Barcelona, Spain}
\date{\today}
\begin{abstract}
Spin-dependent features in the conductivity of graphene, chemically modified by a random distribution of hydrogen adatoms, are explored theoretically. The spin effects are taken into account using a mean-field self-consistent Hubbard model 
derived from first-principles calculations. A Kubo-Greenwood transport methodology is used to compute the spin-dependent transport fingerprints of weakly hydrogenated graphene-based systems with realistic sizes. Conductivity responses 
are obtained for paramagnetic, antiferromagnetic, or ferromagnetic macroscopic states, constructed from 
the mean-field solutions obtained for small graphene supercells. Magnetoresistance signals up to $\sim 7\%$ are calculated 
for hydrogen densities around $0.25\%$. These theoretical results could serve as guidance for experimental observation of induced magnetism in graphene.
\end{abstract}

\maketitle

{\it Introduction}.- Close to the charge neutrality point, two-dimensional graphene 
exhibits fascinating transport properties deriving from the massless Dirac fermions 
physics~\cite{Geim,Castro}. These electronic properties can be further tuned and diversified
 through chemical functionalization~\cite{LKP} or irradiation (defects 
formation)~\cite{Krasheninnikov}. Indeed, strong modifications of ${\rm sp}^{2}$-bonded 
carbon network by grafting ${\rm sp}^{3}$-type defects is, for instance, an interesting strategy 
to induce an insulating behavior in graphene. Following this idea, recent attempts based 
on hydrogenation and fluorination were able to turn graphene into a wide band-gap 
insulator~\cite{Elias,JZhu2010}. 

While ideal bidimensional graphene is nonmagnetic, its unidimensional derivatives (graphene nanoribbons) 
could exhibit magnetic ordered states due to the presence of edges~\cite{Rossier07}. Additionally, monovacancies
 or ${\rm sp}^{3}$-type defects (such as those created by adsorbed atomic H) can introduce local sublattice imbalances in
 graphene, thus inducing magnetic moments~\cite{Yazyev,Palacios2008} according to Lieb's theorem~\cite{Lieb1989}. Experimentally, 
the existence of intrinsic magnetism in graphene and graphite-based materials is strongly debated~\cite{Ugeda,Kats}.
Some experimental works assume that the observed magnetic ordering is coming from graphene edges or grain boundaries~\cite{Kats}, a scenario that has 
been recently questioned by Sepioni and coworkers who found no magnetism even for high concentration of edge defects~\cite{Sepioni}. 
Differently, in a recent experiment~\cite{RTH}, signatures of room temperature ferromagnetism in metal free untreated graphite were tentatively related 
to the presence of hydrogen impurities. We note that in all those experiments, the nature and quantity of defects remain poorly 
characterized and mostly uncontrolled. It would be consequently highly desirable to clarify the conditions for observing spin-dependent transport 
fingerprints in hydrogenated graphene. 

In this Letter, spin-dependent transport properties of hydrogenated graphene-based systems with realistic sizes are explored theoretically. 
Firstly, a spin-dependent one-orbital Hubbard Hamiltonian, describing adsorbed hydrogen impurities on a graphene sheet, is derived using first-principles
 calculations and solved at a mean-field level. The local spin texture around hydrogen adatoms is found to be strongly dependent on both
the concentration and the specific adsorption sublattice of neighboring adatoms as previously
reported~\cite{Palacios2008,Soriano2010}. The self-consistent Hubbard Hamiltonian is then implemented into a real space (Kubo)-transport method. 
Our theoretical results reveal the suitable hydrogenation coverage of graphene to maximize 
spin-dependent conductivities, which translate into measurable magnetoresistance signals in the diffusive regime.

{\it One-orbital mean-field Hubbard approximation}.- The one-orbital {\it tight-binding} 
description of graphene restricts to $p_z$ orbitals centered on each carbon atom.  
The spin-related physics induced by Coulomb interaction is introduced in the calculation by means of the Hubbard
 model in its mean-field approximation 
\begin{equation}
\label{MFHM}
\mathcal{H} = t\sum_\mathrm{<i,j>,\sigma} c_\mathrm{i,\sigma}^\dagger c_\mathrm{j,\sigma} + 
U\sum_\mathrm{i} n_\mathrm{i,\uparrow} \langle n_\mathrm{i,\downarrow} \rangle +
 n_\mathrm{i,\downarrow} \langle n_\mathrm{i,\uparrow} \rangle,
\end{equation} 
%
%
\noindent where $t$ is the first-neighbors hopping term, $c_\mathrm{i,\sigma}^\dagger$ ($c_\mathrm{j,\sigma}$) is the creation (annihilation) operator in the lattice site $i$ ($j$) with spin $\sigma$,  $U$ is the on-site Coulomb repulsion, and $n_\mathrm{i,\downarrow}$, $n_\mathrm{i,\uparrow}$ are the self-consistent
occupation numbers for spin-down and spin-up electrons, respectively. The ratio $U/t$ has been chosen to accurately
 reproduce various physical magnitudes extracted from first-principles calculations (as explained below).
\begin{figure}[h!]
\begin{center}
\includegraphics[width=0.85\linewidth]{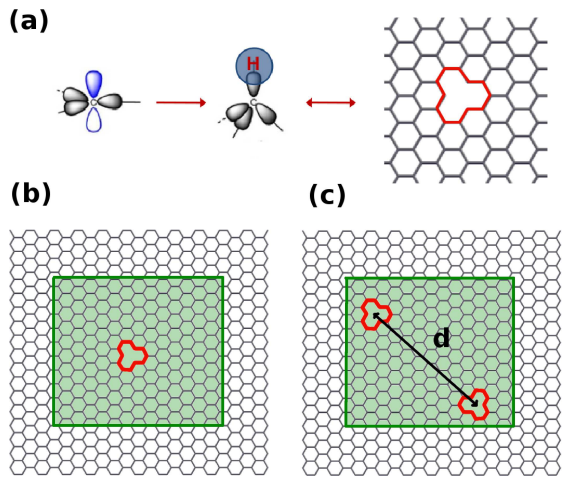}
\caption{(color online) (a) Schematic representation of the $sp^2$ hybridization 
breakdown upon hydrogen adsorption. In the present one-orbital {\it tight-binding} 
model, such a defect is equivalent to the creation of a monovacancy. 
(b) Single-vacancy supercell used in the mean-field Hubbard model calculations with periodic boundary conditions. (c) Supercell with two vacancies. The distance between two hydrogen defects is labeled by $d$. The green colored region give the space location for the defects, to minimize the correlations between vacancies in neighboring cells.}
\label{Fig1}
\end{center}
\end{figure}
When a hydrogen (H) atom is adsorbed on top of a carbon (C) atom, the $sp^2$-symmetry is locally broken, and the electron from the C $p_z$ orbital is removed from the $\pi$ bands to form a $\sigma$ bond with the H atom. Consequently, within the 
{\it tight-binding} approximation, such an  adsorption is modeled by removing the 
corresponding electron and lattice site [as illustrated in Fig.\ref{Fig1} (a)]. Figure~\ref{Fig1}(b) presents one of the supercells 
($N=791$ atomic sites) used in periodic boundary conditions calculations. This hydrogen defect induces zero-energy
 electronic states which are mainly localized around the impurity~\cite{Ugeda,CastroNeto}. When the Coulomb repulsion is switched on 
in the calculation, these zero-energy states spin-polarize, leading to 
semi-local $S=1/2$ magnetic moments with a staggered spin density mostly located on one 
sublattice~\cite{Yazyev,Palacios2008,Soriano2010} [see Fig.\ref{Fig2} (a,d)]. 
When a finite concentration of defects is present, the magnetic ordering 
between these magnetic moments is dictated by the
sublattices where they are located, being co-polarized or ferromagnetic (FM) for the same sublattice
and counter-polarized or antiferromagnetic (AF) otherwise [Fig.\ref{Fig2}(b,e)].
\\
\\
{\it Magnetic states}.-
The total spin $S$ of the macroscopic ground state is given by 
the excess of magnetic moments on one specific sublattice, 
although $S=0$ is the most likely value on simple statistical grounds
(equal H occupation of both sublattices).
In the following, two assumptions related to the macroscopic magnetic 
states are considered. The first one is the simplest: a \textit{paramagnetic} (PM) state, which is a good approximation for sufficiently low H concentrations, or at sufficiently high temperatures, in which there is
no correlation between the semi-local moments.
This state is thus constructed from independent large supercells containing a 
single vacancy as shown in Fig.\ref{Fig1} (b). The corresponding spin densities
on each cell site is represented in Fig.\ref{Fig2}(a,d) as a function of the distance to the center of the supercell
for two different concentrations of defects. The result favorably compare with first principles calculations when $U/t$ 
is chosen appropriately~\cite{ABINITIO}. 

%
\begin{figure}[h!]
\begin{center}
\includegraphics[width=0.95\linewidth]{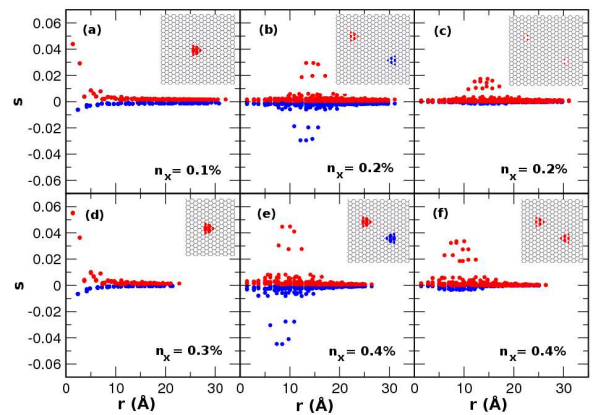}
\caption{(color online) Local (site) spin ($s$) versus $r$ (the distance to the center of the supercell) 
for various vacancy concentrations: (a,d) $n_x=0.1\%$ and $0.3\%$ in the paramagnetic case, (b,e) $n_x=0.2\%$ ($d=27.7 \AA$) and $0.4\%$ ($d=18.1 \AA$) in an antiferromagnetic state, and (c,f) in a ferromagnetic
 excited state within the mean-field Hubbard approximation [$\uparrow$ 
(red) and $\downarrow$ (blue) electrons]. In all cases, the center of mass of the
vacancies is at $r=0$.}
\label{Fig2}
\end{center}
\end{figure}

The second assumption
is more complex since pairwise magnetic ordered states are 
investigated. The latter are constructed using supercells containing two vacancies on different sublattices 
[see Fig.\ref{Fig1} (c)] far from the edges. 
When selecting one site on each sub-lattice [see Fig.\ref{Fig1}(b,e)], the magnetic moments couple antiferromagnetically and the macroscopic state generated by merging several supercells is an antiferromagnet (AF) with zero total spin ($S=0$).  
We explore the effect of the distance ($d$) between vacancies while varying accordingly the supercell size. By changing 
simultaneously the distance and size, 
we always allow for a meaningful definition for the concentration $n_x$ based on only two vacancies per cell. One notes that for $n_x = 0.6\%$ ($d=8.5 \AA$) the vacancies become very close to each other, thus allowing overlapping between magnetic moments and suppression of the local antiferromagnetic state~\cite{Palacios2008}. We note that, 
when the vacancies are in the same sublattice, the ground state is ferromagnetic (FM), but we will
not consider this case in this work. 

The AF ordering can be further tuned by applying a sufficiently large external magnetic field. This has been estimated to lie in the range $B \approx 1-10$ T for hydrogenated graphene nanoribbons\cite{Soriano2010} and similar distances between H adatoms. Since the magnitude of the local magnetic moments is lower in the case of graphene due to the absence of confinement, we expect the number to be smaller in the present case. On application of the field, the local magnetic moments become co-polarized giving rise to a FM state. The difference in the magnetic ordering can affect the conductivity and induce a magnetoresistive (MR) response as discussed for graphene nanoribbons in Ref.~\onlinecite{Soriano2010}. To study the possible magnetoresistance signals in hydrogenated graphene samples, the spin density and related spin-dependent on-site energies have been computed for this specific case [Fig.\ref{Fig2}(c,f)]. 
\\
\\
{\it Kubo conductivity methodology}.- Using the self-consistent Hubbard Hamiltonian calculations, spin-dependent 
on-site energies are estimated for each supercell. These {\it tight-binding} parameters are then used 
for computing the conductivity in realistic-sized samples applying the Kubo formalism. These graphene samples are constructed 
by attaching a large number of these supercells following different random arrangements. An efficient
 real space order N method is then used to simultaneously follow the wavepackets dynamics and to 
compute the Kubo conductivity (see Refs.~\onlinecite{Kubo1} for details).  The analysis of the wavepacket dynamics
 allows to extract the conduction mechanisms and transport length scales (mean free path). As long as $n_{x}\leq 0.5\%$ the diffusion coefficients are found to reach a saturation regime after a few thousands of femtoseconds, and then remain 
roughly constant up to very long times, indicating a diffusive regime and a negligible contribution of 
quantum interferences within the computational reach. The maximum value allows the evaluation of $\ell_{e}=D_{max}/v_{F}$, with 
$v_{F}=10^{6}{\rm ms^{-1}}$.  Note that the transport properties are calculated in the zero temperature limit, which
should however be robust to higher temperatures given the weak electron-phonon coupling in graphene.
\begin{figure}[h!]
\begin{center}
\includegraphics[width=\linewidth]{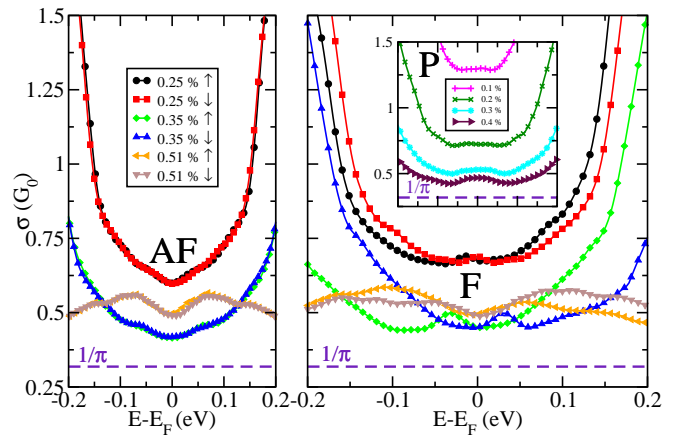}
\caption{(color online) Left panel : Conductivities of hydrogenated graphene samples with various
concentrations for the antiferromagnetic case (left panel), paramagnetic case 
(right panel inset) and ferromagnetic case (right panel main frame). 
Dashed lines give $2e^{2}/\pi h$.}
\label{Fig3}
\end{center}
\end{figure}
\\
\\
{\it Spin-dependent transport features}.- The scaling analysis of the conductivity is carried out using the Kubo 
formula $\sigma_{\uparrow,\downarrow} (E,t)=(e^{2}/2) \hbox{Tr}[\delta_{\uparrow,\downarrow} 
(E-\hat{H})]D_{\uparrow,\downarrow}(E,t)$ where $\text{Tr}[\delta_{\uparrow,\downarrow} (E-\hat{H})/S]$ are the
 spin-dependent DoS per unit of surface at the energy E. At a fixed energy, $\sigma_{\uparrow,\downarrow}(E,t)$ 
reach their maximal values at the same time as $D_{\uparrow,\downarrow}(E,t)$ ($D_{\uparrow,\downarrow}^\text{max}$) which gives the
 spin-dependent Drude conductivities $\sigma^{\rm Drude}_{\uparrow,\downarrow}(E)=(e^{2}/2) 
\hbox{Tr}[\delta_{\uparrow,\downarrow}(E-\hat{H})]D_{\uparrow,\downarrow} ^\text{max}(E)$), that accounts for 
the disorder effects on the density of states.
\\
\\
Fig.~\ref{Fig3} presents the Drude conductivities for the AF (left panel),  
FM cases (right panel) and the approximated PM case (inset in right panel) for H concentrations varying from $0.1\%$ up to $0.5\%$. 
As expected, both in the AF and the PM case, no spin-dependent transport features
are observed in the results, with the conductivities for up and down spins being
equal within the statistical error.  We note, however, some differences in the shape of 
the conductivity curves versus energy close to the Dirac point between the PM and AF cases. 

\begin{figure}[htbp]
\begin{center}
\includegraphics[width=\linewidth]{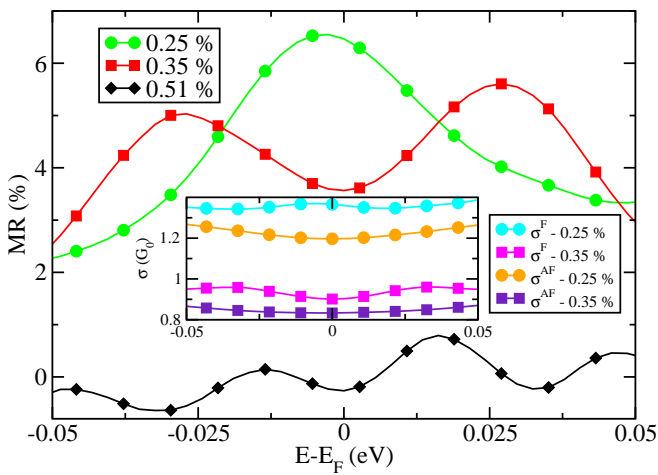}
\caption{(color online) Magnetoresistance of hydrogenated graphene samples with concentrations of 
$n_{x}=0.25,0.35,0.5\%$ as a function of charge energy. }
\label{Fig4}
\end{center}
\end{figure}

In sharp contrast, in the FM case, the spin-dependent conductivities significantly differ away from the charge neutrality point, with $|\sigma^{\rm Drude}_{\uparrow}(E)/ \sigma^{\rm Drude}_{\downarrow}(E)|$ typically increasing with energy.  $\sigma^{\rm Drude}_{\uparrow,\downarrow}(E)$ decays roughly linearly with $n_{x}$ for the lowest H density (which is a trend general for resonant scattering~\cite{Wehling}), but one clearly observes a tendency towards saturation at higher concentration. 
It is worth mentioning that, in all cases, the total Drude conductivities ($\sigma^{\rm Drude}_{\uparrow}(E)+\sigma^{\rm Drude}_{\downarrow}(E)$) 
remain larger than $2G_{0}/\pi=4e^{2}/\pi h$, which is the theoretical minimum value estimated within the self-consistent Born 
approximation~\cite{Ando}.
\\
\\
>From the computed conductivities in the AF and FM cases, one extracts (for AB-sublattice 
preserved symmetry), the magnetoresistance $\text{MR} = (\sigma^{F}-\sigma^{AF})/(\sigma^{F}+\sigma^{AF})$, 
with $\sigma^{F} = \sigma^{F}_{\uparrow} + \sigma^{F}_{\downarrow}$ and
$\sigma^{AF} = \sigma^{AF}_{\uparrow} + \sigma^{AF}_{\downarrow}$. Fig.~\ref{Fig4} shows MR for three different H concentrations. The maximum signal is observed for the lowest density $n_{x}=0.25\%$ and reaches about $7\%$ for a switch from the antiferromagnetic to the excited ferromagnetic case. The signal is reduced with increasing $n_{x}$ and tends to disappear at higher concentrations, along with the spin density. One also notes that, since the macroscopic states are built out of supercells carrying only two vacancies, the number of magnetic correlations is likely underestimated. Hence, experimental observations of larger MR values would not be a surprise. We observe some asymmetry in the MR profiles (Fig.4) which roots in small numerical inaccuracies stemming from both, the approximated Hubbard Hamiltonian and the use of a random phase approximation for the Kubo methodology. Fig.4 (inset) also emphasizes the energy-dependence conductivities for $n_{x}=0.25\%$ and $n_{x}=0.35\%$ close to Fermi level, which result in the qualitatively different profiles observed in the energy-dependent MR signals.
\\
\\
{\it Conclusion}.-Spin-dependent conductivities in disordered, hydrogenated graphene have been computed and related to the underlying magnetic ordering taking place at a large scale. Magnetoresistance signals of up to $7\%$ have been found in the low density limit, suggesting that the existence of local magnetic states in graphene, induced by hydrogenation, could be observed experimentally. The high concentration limit ($n_{x}\geq 1\%$) in which stronger quantum interferences and localization effects should take place deserves a further consideration. 
\\
\\
{\it Acknowledgements}.-Funding by Spanish MICINN is acknowledged by D.S. and J.J.P. (Grants  FIS2010-21883-C02-02 and CSD2007-00010) and P.O. (Grants FIS2009-12721-C04-01 and CSD2007-00050). D.S. was supported by the PhD Grant Program from CSIC and the Unidad Asociada of the Universidad de Alicante and acknowledges Joaqu\'\i n Fern\'andez-Rossier for discussions.  This work is connected to the Belgian Program on Interuniversity Attraction Poles (PAI6), to the NanoHymo ARC sponsored by the Communaut\'e Fran\c{c}aise de Belgique, to the ETSF e-I3 project (Grant ${\rm N.}^{\circ}$ 211956), and to the NANOSIM-GRAPHENE Project 
${\rm N.}^{\circ}$ ANR-09-NANO-016-01. J.-C.C. acknowledges funding from the FNRS of Belgium. Computational resources were provided by the CISM of the Universit\'e Catholique de Louvain. 


\begin{thebibliography}{50}
\bibitem{PMS} Due to the zero-energy
wave-function decay close to a hydrogen adatom in graphene, $\Psi \propto 1/\mathbf{r}$, the supercell size
 must be chosen large enough to avoid interaction between magnetic moment localized around the defect and its periodic images 
in the neighboring cells. This convergence study is performed by checking that on-site energies tend to $U/2$ and that the spin density 
approaches zero near the border of the supercells. Although the spin density never truly vanishes near the border, its value can be neglected 
for the considered vacancy concentrations.
\end{thebibliography}
\end{document}